\documentclass[showpacs,preprintnumbers,amsmath,amssymb,twocolumn,prb]{revtex4}

\usepackage{graphicx}
\usepackage{times}
\usepackage{amsmath}
\usepackage{amssymb}
\usepackage{bm}         % bold math
\usepackage{color}
\usepackage[utf8]{inputenc}
\usepackage[T1]{fontenc}
\usepackage{mathptmx}

\begin{document}

\title{Energy spectra of graphene quantum dots induced between Landau levels}

\author{G. Giavaras}
\affiliation{CEMS, RIKEN, Wako-shi, Saitama 351-0198, Japan}

\pacs{73.21.La,73.23.-b,81.05.ue}

\begin{abstract}
When an energy gap is induced in monolayer graphene the valley
degeneracy is broken and the energy spectrum of a confined system
such as a quantum dot, becomes rather complex exhibiting many
irregular level crossings and small energy spacings which are very
sensitive to the applied magnetic field. Here we study the energy
spectrum of a graphene quantum dot that is formed between Landau
levels, and show that for the appropriate potential well the dot
energy spectrum in the first Landau gap can have a simple pattern
with energies coming from one of the two valleys only. This part
of the spectrum has no crossings, has specific angular momentum
numbers, and the energy spacing can be large enough, consequently,
it can be probed with standard spectroscopic techniques. The
magnetic field dependence of the dot levels as well as the effect
of the mass-induced energy gap are examined, and some regimes
leading to a controllable quantum dot are specified. At high
magnetic fields and negative angular momentum a simple approximate
method to the Dirac equation is developed which gives further
insight into the physics. The approximate energies exhibit the
correct trends and agree well with the exact energies.
\end{abstract}

\maketitle

\section{Introduction}

Some experiments~\cite{freitag16, freitag18, li, lee} have shown
that a quantum dot can be formed in a single sheet of graphene by
adjusting the tip-induced potential of a scanning tunnelling
microscope (STM). In a uniform magnetic field discrete energy
levels have been probed between bulk Landau
levels,~\cite{freitag16, freitag18} and the electronic properties
of confined states have been explored.~\cite{freitag16, freitag18,
li, lee, qiao} To better control the tunability of the quantum
devices in the experimental studies an energy gap is usually
induced at the Dirac points $K$ and $K'$ of the band structure.
Without the energy gap the two valleys at the Dirac points are
degenerate and the charge carriers are massless.~\cite{rozhkov}
Inducing an energy gap between the conduction and the valence
bands leads to charge carriers with mass and the valley degeneracy
is broken.

In a graphene sheet the energy gap can be induced with some simple
techniques~\cite{balog, wang, zhou, fan, shemella} however,
breaking the valley degeneracy leads to more complex energy
spectra for a confined graphene system, since the energies of the
charge carriers depend on the specific valley. The resulting
energy spectra exhibit many level crossings as well as
anti-crossings, and small energy spacings which are rather
sensitive to the applied magnetic field. These features could make
difficult the use of graphene dots to opto-electronics and
valleytronics applications.~\cite{schaibley, karanikolas}

In the present work we study the energy spectrum of a graphene
quantum dot that is formed between bulk Landau levels, and show
that for the appropriate fields the dot energy spectrum can have a
simple pattern. Specifically, if we focus on the first Landau gap,
namely, between the Landau levels $-1$ and $0$, then there is an
energy range with discrete energies coming from one of the two
valleys only. Thus, a graphene dot with a specific valley index
can be realized. We demonstrate that the dot energy spectrum
consists of specific angular momentum values and has no crossings,
simplifying drastically the identification of the dot energy
levels. Our calculations show that when the mass term is a few
tens of meV and the applied magnetic field is a few Tesla, the
typical energy spacing can be a few meV. By adjusting the STM
induced potential well the discrete levels of the dot formed in
the first Landau gap can be energetically isolated and lie away
from the bulk Landau levels.

To obtain further insight into the physics we develop an
approximate method to compute the dot energies. The method takes
into account the fact that the zeroth Landau level has a Dirac
state with one component zero, and even though this component
becomes nonzero in the presence of the STM potential, it can still
be vanishingly small compared to the second component. Using this
condition we derive an approximate Schrodinger equation for the
dominant component of the Dirac state. We find that at high fields
and negative angular momentum the approximate energies exhibit the
correct potential dependence, and are in a good agreement with the
exact ones. The typical error is of the order of $1\%-2\%$, and
depending on the specific parameters the error can decrease to
less than $0.1\%$.

Only in the proper range of parameters the quantum dot energy
spectra have a simple pattern. Anticrossing points which are
relevant to the Klein tunnelling effect do not occur in the range
of parameters used in this work. Quantum dots at low magnetic
fields, typically lower than $0.5$ T, exhibiting Klein tunnelling
have been theoretically examined earlier~\cite{giavaras09,
giavaras12} but these dots might be more difficult to probe due to
the relatively small energy spacings and the fabrication of the
required dot confining potential.~\cite{giavaras09} Quantum dots
in the Klein tunnelling regime have also more complicated quantum
states and the approximate method developed in this work is
inapplicable.

The quantum dot system studied here is edge-free therefore the
reported results are insensitive to the microscopic details of the
edges. This property offers a superior control over the dot states
compared to dot states formed in nano-sheets of
graphene.~\cite{rozhkov} The results are relevant to other
confined systems formed in two-dimensional materials by external
fields,~\cite{moriyama, abdullah, caneva, keren, recher, jakubsky,
sadrara, giavaras11, kandemir, maksym} as well as to general
hybrid graphene-based systems including defects and
heterostructures.~\cite{gold, yuli, walkup, giorgos, schattauer,
mirzakhani, myli, dnle}

Section~\ref{model} presents the physical model of the graphene
quantum dot and explains with some qualitative arguments how the
discrete levels of the dot emerge from the bulk Landau levels. The
energy range of interest is also specified in Sec.~\ref{model},
and then in Sec.~\ref{levels} the energy spectra of the quantum
dot are studied. In Sec.~\ref{approxi} an approximate method to
derive the dot energies is presented and a comparison with the
exact energies is made. The conclusions are presented in
Sec.~\ref{conclu}. Finally, in Appendices A, B and C the quantum
dot equations are derived and some further results are presented.

\section{Quantum dots formed in the first Landau gap}\label{model}

In the continuum approximation the charge carriers in graphene
satisfy the Dirac equation~\cite{rozhkov}
\begin{equation}\label{dirac}
[v_{\text{F}}\bm{\sigma} \cdot ({\bf p} + e {\bf A})+ V +
\tau\Delta\sigma_z/2 ]\Psi =E \Psi,
\end{equation}
with $v_{\text{F}}=10^{6}$ m s$^{-1}$, ${\bf p} = (p_x, p_y)$ is
the momentum operator, and $\bm{\sigma}=(\sigma_x, \sigma_y)$,
$\sigma_z$ are the Pauli matrices. A uniform magnetic field $B$ is
perpendicular to the graphene sheet in the $z$-direction and the
magnetic vector potential is $A_{\theta}(r)= B r /2$ in the
azimuthal direction, where $r$ is the radial coordinate. The
scalar potential has the form $V(r)=-V_0\exp(-r^2/2L^2)$ which
models the STM-induced potential well with an effective depth
$V_0\ge0$. The effective width is controlled by the parameter $L$
which is taken to be $\sqrt{2}L=40$ nm unless otherwise specified.
The mass term is denoted by $\Delta/2$ giving rise to an energy
gap equal to $\Delta$, and $\tau=1$ $(-1)$ denotes the $K$ ($K'$)
valley. Because of the cylindrical symmetry the two-component
envelope function can be written as~\cite{giavaras12, giavaras11}
$\Psi(r, \theta) = (f_{1}(r)\exp[i(m-1)\theta]$, $if_{2}(r)\exp[i
m\theta])/\sqrt{r}$, where $m=0,\pm 1,...$ is the angular momentum
quantum number, and $\theta$ is the azimuthal angle. Here,
$f^{2}_{i}$ is the radial probability distribution for each of the
two sublattices of the graphene sheet. Equation~(\ref{dirac}) is
discretized with a finite-difference scheme~\cite{giavaras09}
applying the appropriate boundary conditions for confined states
that satisfy $E(m, B, \tau) = E(1-m, -B, -\tau)$. This scheme
converts the continuum eigenvalue problem Eq.~(\ref{dirac}) to a
matrix eigenvalue problem which is solved numerically.

When there is no potential, $V_0=0$, Eq.~(\ref{dirac}) gives rise
to the Landau levels. Defining the Landau index $N = n + (|m| + m
)/2$ with the integer $n=0, 1, \dots$ then the Landau levels for
$N=0$ ($n=0$ and $m\le0$) are $E =  - \tau \Delta/2$. The excited
Landau levels for $N\geq 1$ are $E = \pm \sqrt{ 2 \hbar
v^{2}_{\text{F}} N e B+ \Delta^2/4}$.

When $\Delta=0$ all $m\le0$ give a Landau energy $E=0$ independent
of the valley. On the contrary, when $\Delta\ne0$ all $m\le0$ for
$\tau=-1$ give a Landau energy $E=\Delta/2$, and all $m\le0$ for
$\tau=1$ give a Landau energy $E=-\Delta/2$. As quantified below,
by increasing $V_0\ne0$ the original Landau energies
$E=\pm\Delta/2$ start to decrease with a rate that depends on
$|m|$; small $|m|$ energies are affected more by $V_0$. Thus, the
energies $E=\Delta/2$ form a set of discrete energies with
$\tau=-1$, and similarly the energies $E=-\Delta/2$ form another
set of discrete energies with $\tau=1$. When the Landau gaps are
large enough only these two energy sets are relevant and the
typical energy separation between them is of the order of
$\Delta$, while the two sets overlap when $V_0\gg\Delta$. The key
conclusion is that discrete energies which come from different
valleys can lie in a very different energy range, for the proper
parameter regime, allowing the realization of a dot with a
well-defined valley. All these arguments are quantified below, and
it is also shown that the typical spacing of the discrete energies
lying in the first Landau gap is large enough as needed to define
a controllable quantum dot. In this work the first Landau gap,
which is denoted by $E_L$, is defined from $-\sqrt{ 2 \hbar
v^{2}_{\text{F}} e B+ \Delta^2/4}$ to $-\Delta/2$, and this energy
gap specifies the energy range of interest.

For $V_0\ne0$ a numerical approach is needed to compute the
energies, however, some insight into the valley-dependent energy
range can be obtained without solving Eq.~(1). If we set $f_i=y_i
g_i$ with $i=1$, 2 and $g_i$ is determined explicitly in Appendix
A, then the function $y_i$ satisfies the second order differential
equation $d^{2}y_i/dr^2_{i}+Q^{2}_{i}y_{i}=0$. The energy
dependent coefficient is
\begin{equation}\label{Q2}
\begin{split}
Q^{2}_{i}(r, E) & = \mp \frac{U'}{\gamma} \pm \frac{U}{\gamma}
\frac{V'}{V_{\mp}} + \frac{V^{''}}{2V_{\mp}}-
\frac{3}{4}\left(\frac{V^{'}}{V_{\mp}}\right)^2-\frac{U^2}{\gamma^2}\\
&+ \frac{(V-E)^2}{\gamma^2} -\frac{\Delta^2}{4\gamma^2},
\end{split}
\end{equation}
with the parameter $\gamma = v_{\text{F}}\hbar$. The upper/lower
sign is for $i=1/2$, $V_{\mp} = V-E\mp\tau\Delta/2$,
$U/\gamma=(2m-1)/2r + eA_{\theta}/\hbar$, and prime denotes
differentiation with respect to $r$.

Unlike the $V_0=0$ limit for $V_0\ne0$ a confined state $\Psi$ can
exist when both components $y_i$ are non-zero. However, this
condition is guaranteed only for specific forms of $Q^{2}_{i}(r,
E)$. As an example, consider the parameters $B=1$ T, $\Delta=20$
meV, $V_0=20$ meV and for brevity focus on $m=-1$ and energies in
the first Landau gap. A non-zero $y_i$ can exist when there is a
spatial region with $Q^{2}_{i}(r)>0$; $y_i$ is localised in this
region and $Q^{2}_{i}(r)<0$ asymptotically and for
$r\rightarrow0$. A non-zero $y_i$ can also exist when
$Q^{2}_{i}(r)<0$ for all $r$ but $R_{0}$ with $V_{\mp}(R_0)=0$
resulting in $Q^{2}_{i}(R_0)\rightarrow-\infty$. Then, $y_i$ is
localised in the vicinity of $R_0$. However, for $\tau=-1$ no
energies satisfy the required forms for both coefficients
$Q^{2}_{i}(r)$, and simultaneously the limit $E \rightarrow
\Delta/2$ as $V_0 \rightarrow 0$. In contrast, for $\tau=1$ and
for specific energies both $Q^{2}_{i}(r)$ have the required forms.
The exact energies for the $\tau=1$ valley can computed by solving
the equations for $y_{i}$. Increasing now $V_0$ to $V_0=72$ meV,
and exploring again the form of $Q^{2}_{i}(r)$ in the first Landau
gap shows that both $y_i$ can be non-zero for both valleys
$\tau=\pm1$. This example demonstrates that with the proper choice
of $V_0$, which can be controlled with the STM tip-induced
potential, energies from the $\tau=1$ valley only, or from both
valleys can be formed in the first Landau gap.

\begin{figure}[t]
%%% Requires \usepackage{graphicx}
\includegraphics[width=5.9cm, angle=270]{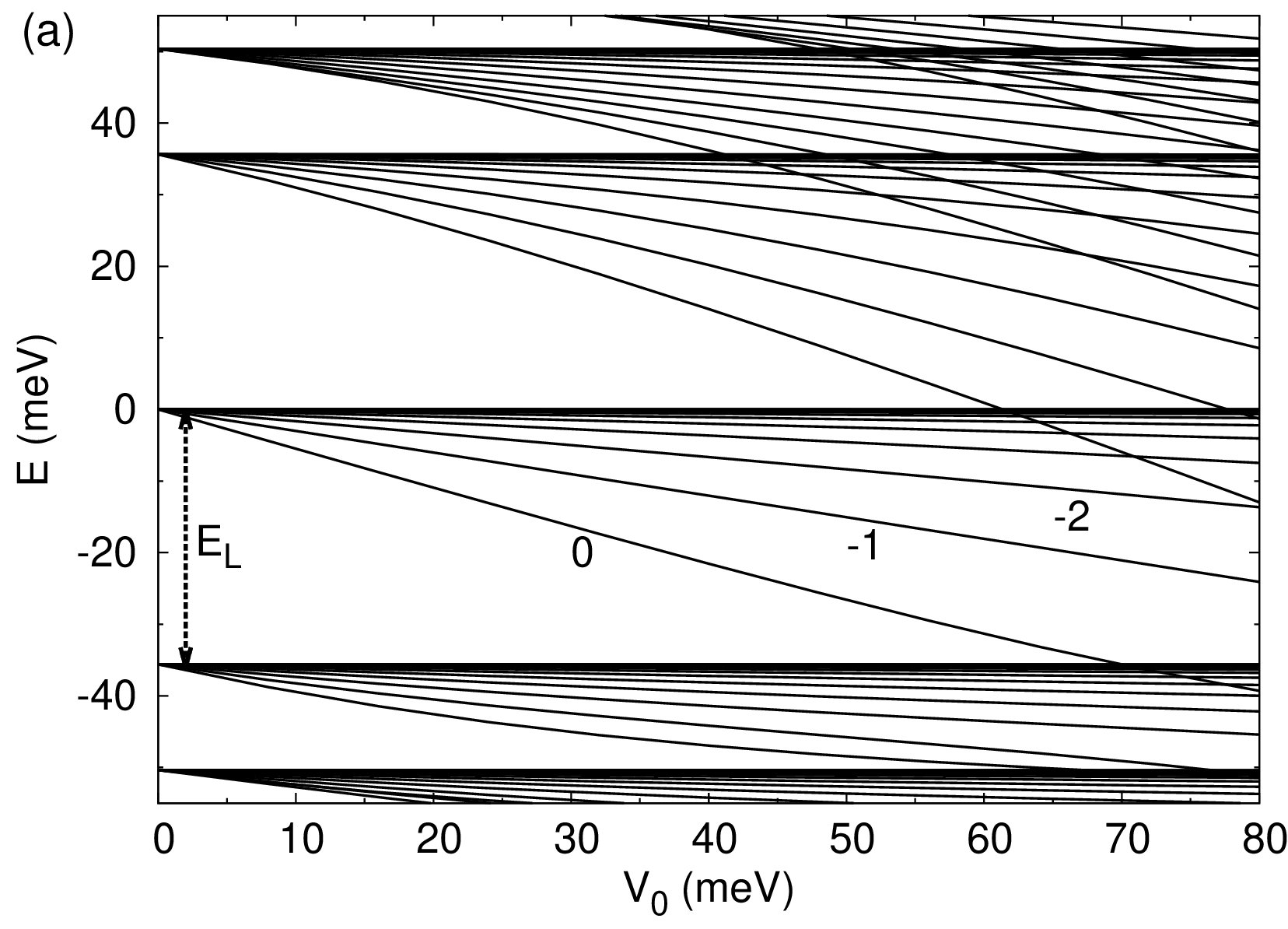}
\includegraphics[width=5.9cm, angle=270]{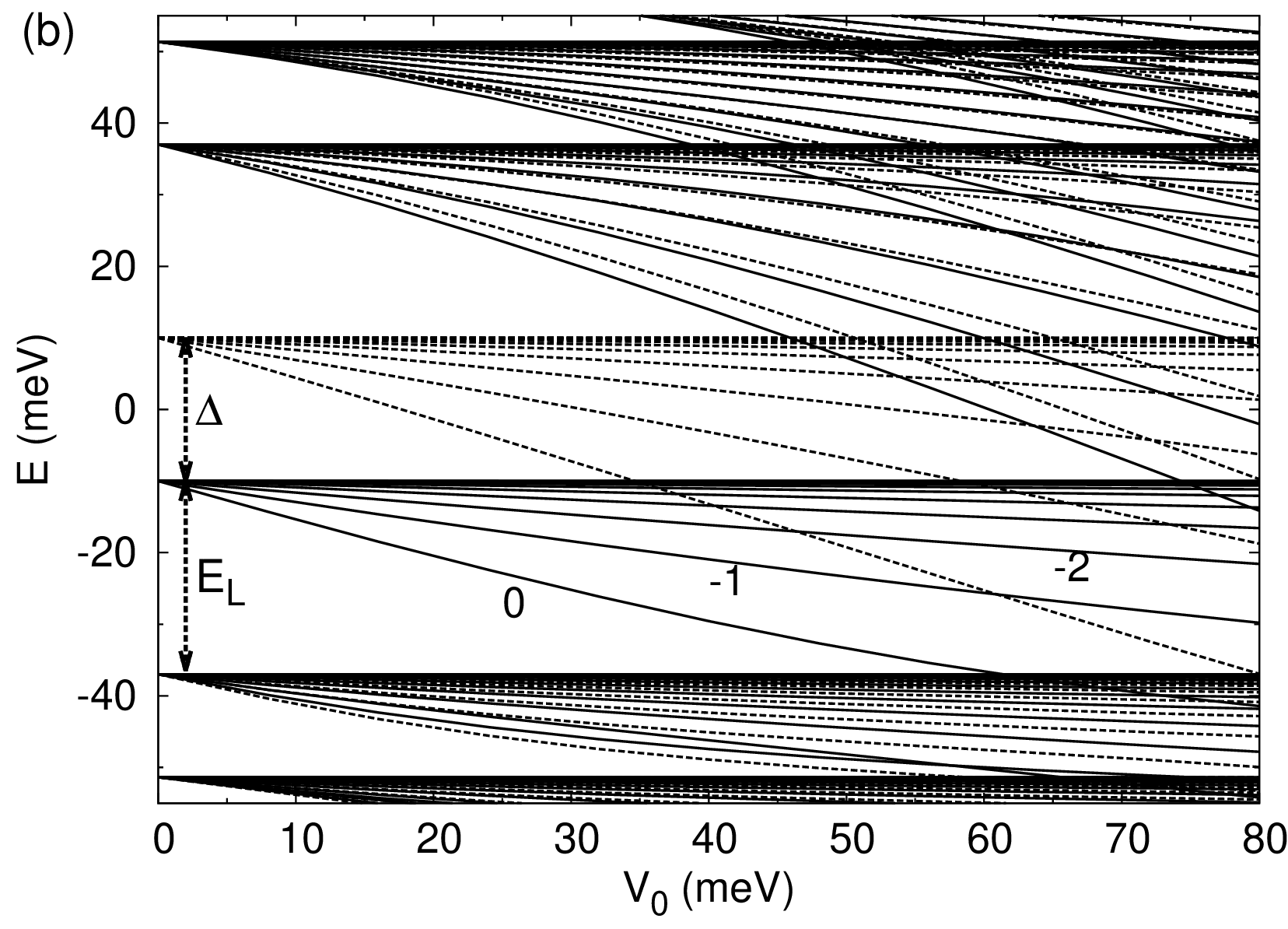}
\includegraphics[width=5.9cm, angle=270]{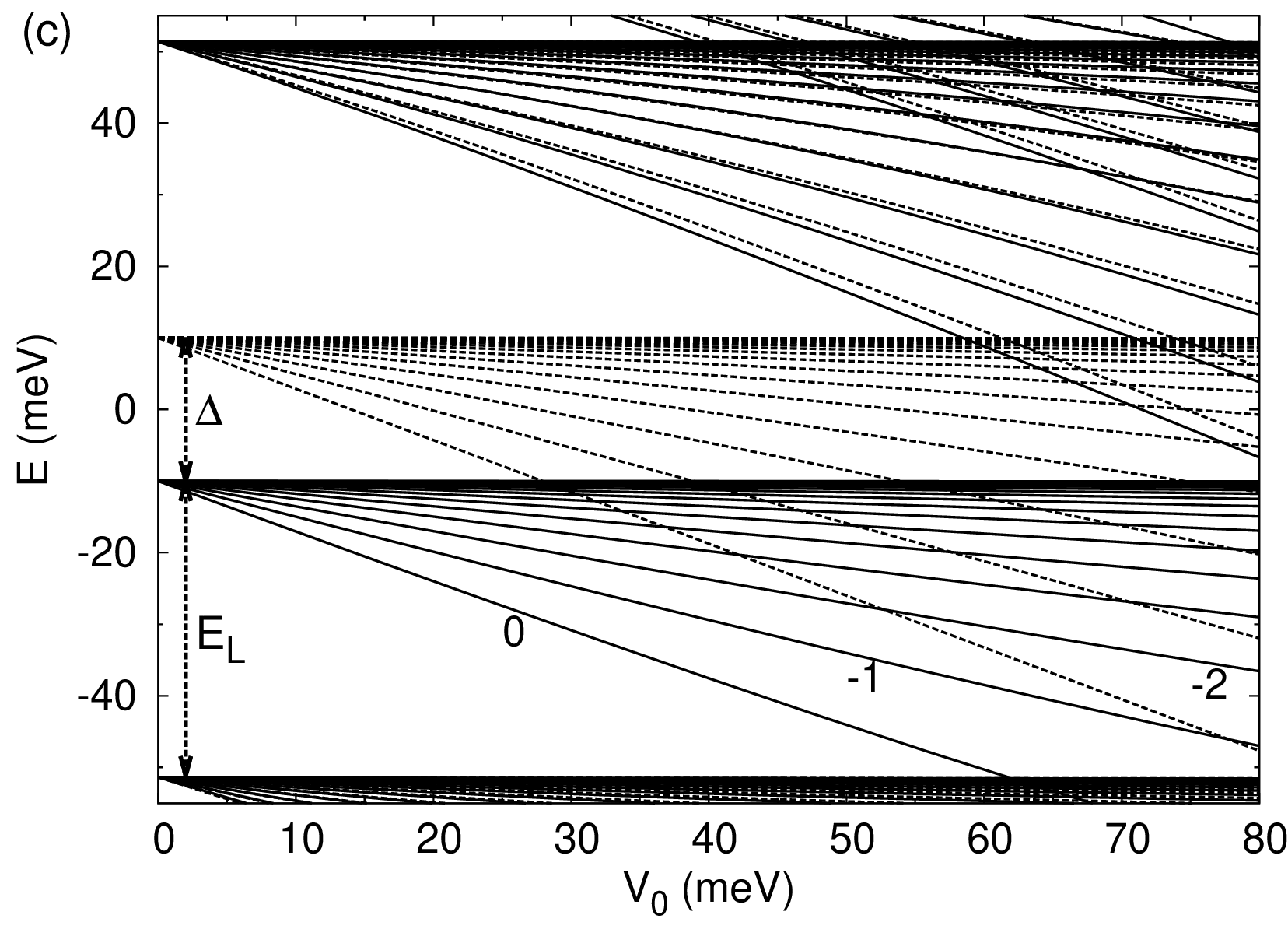}
\caption{Energies as a function of the potential well depth for
the parameters: (a) $B=1$ T, $\Delta=0$ (b) $B=1$ T, $\Delta=20$
meV (c) $B=2$ T, $\Delta=20$ meV. Solid lines correspond to
$\tau=1$ and dashed lines to $\tau=-1$. The first Landau gap
relevant to this work is denoted by $E_L$ and the mass-induced gap
by $\Delta$. The numbers 0, -1, -2 denote the angular momentum
number.}\label{spectrum}
\end{figure}

\section{Quantum dot energy spectra}\label{levels}

In this section the energy spectra of the graphene quantum dot are
examined with exact numerical calculations.
Figure~\ref{spectrum}(a) shows the energy levels as a function of
the potential depth $V_0$ for $B=1$ T and $\Delta=0$. For
$V_{0}=0$ the energies define the bulk Landau levels while by
increasing $V_{0}$ discrete energy levels start to form in the
Landau gaps. At a fixed $V_{0}$, the number of energy levels in
each Landau gap is different because the effect of the potential
on the levels depends on the corresponding angular momentum number
$m$. The general rule is that states with large $|m|$ which are
localised away from the potential well are affected less by
changes in $V_{0}$, and their energies deviate only slightly from
the Landau levels. On the contrary, states with small $|m|$ are
localized nearer the origin of the potential well and are more
sensitive to changes in $V_{0}$. As a result, increasing $V_{0}$
shifts the corresponding energies in the Landau gaps by a
significant amount.

In a graphene sheet the bulk Landau levels are broadened due to
impurities, e.g., in the substrate, and/or a disorder potential.
For a controllable quantum dot its discrete energy levels of
interest can be more easily probed when they are defined away from
the Landau levels. Therefore, one possible choice is to define the
dot levels in (the centre of) the first Landau gap which is the
largest. This work is concerned with this case.

Focusing on the first Landau gap $E_L = \sqrt{ 2 \hbar
v^{2}_{\text{F}} e B}$ in Fig.~\ref{spectrum}(a) where $\Delta=0$,
then at $V_{0} \approx 32.5$ meV the lowest energy level is equal
to $E \approx - E_L/2$. This level has $m=0$ and the next higher
level has $m=-1$, then $m=-2$, and so on. Eventually, for large
negative $m$ values the energies are approximately zero defining
the zeroth Landau level $E=0$. The low-lying $m$ energies are well
isolated from other energies and can define the quantum dot
levels. This is not guaranteed when $V_0$ is arbitrary large and
the resulting energy spectrum is more complicated involving both
positive and negative values of $m$ without a specific order. In
Fig.~\ref{spectrum}(a) this trend starts to occur for $V_0>62$
meV. The knowledge of $m$ as well as the dot energy range are
useful since they specify the form of the quantum states, e.g.,
position of peaks and number of nodes.

\begin{figure}[t]
%%% Requires \usepackage{graphicx}
\includegraphics[width=6.3cm, angle=270]{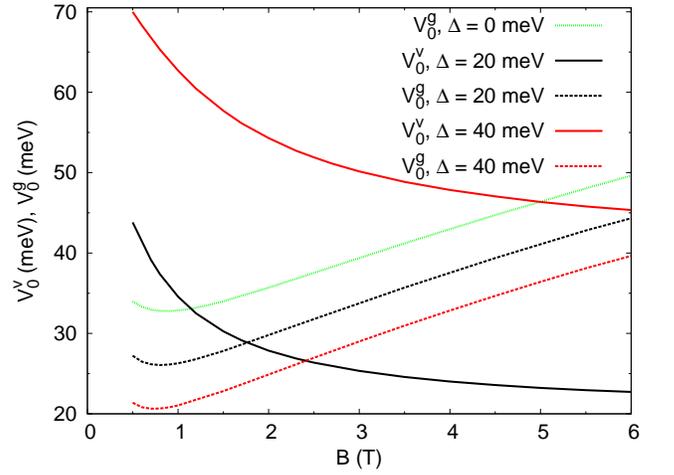}
\caption{Critical potentials $V^{v}_{0}$ and $V^{g}_{0}$ defined
in Eqs.~(\ref{crit1}) and (\ref{crit2}) respectively, as a
function of the magnetic field for $\Delta=0$, 20, 40 meV. For
$V^{v}_{0}> V^{g}_{0}$ the energy levels in the first Landau gap
come from the $K$ valley only, and the lowest energy level in the
first Landau gap lies in the middle of this gap with $m=0$. For
$\Delta=0$ the potential $V^{v}_{0}$ cannot be
defined.}\label{valley}
\end{figure}

In Fig.~\ref{spectrum}(b) the energy levels are plotted for $B=1$
T and $\Delta=20$ meV. Now the energy levels are different for the
two valleys, but the general characteristics of the energy spectra
are similar to those when $\Delta=0$. Discrete energy levels in
the mass-induced gap are induced even when $B=0$ but because of
the symmetry $E(m, \tau) = E(1-m, -\tau)$ the energies from the
two valleys cannot be separated. Furthermore, in a specific device
the mass-induced gap is not so easily tunable as the Landau gap
and its value is rather sensitive to the specific device
configuration. For this reason, the present work is concerned with
the energy levels of a quantum dot formed in the first Landau gap
which is easily tunable by the magnetic field and can be adjusted
at will. In Fig.~\ref{spectrum}(c) the energy levels are plotted
at a higher field, e.g., $B=2$ T but the same $\Delta=20$ meV.
Compared to the $B=1$ T spectrum now more levels, corresponding to
greater values of $|m|$, are introduced in the first Landau gap.
The reason is that by increasing $B$ the states tend to shift
nearer the origin thus the effect of the potential well becomes
more important and the energies start to deviate from the zeroth
Landau level.

For the calculations of the energy spectra the range of angular
momentum numbers $m$ is large enough to accurately derive all
energies in the considered energy range. The convergence of the
energy spectrum for larger values of $V_0$ and $B$ requires a
broader $m$ range. For example, at $V_0=30$ meV and $B=1$ T the
$m=-10$ energy level is about 41 $\mu$eV below the zeroth Landau
level. When the field increases to $B=2$ T then to a good
approximation the same energy difference is observed for $m=-18$.

Figures~\ref{spectrum}(b) and (c) demonstrate that an interesting
energy pattern arises in the first Landau gap provided $V_0$ is
not too large; for example in Fig.~\ref{spectrum}(b) for $V_0 <
35$ meV only energy levels from the $K$ valley ($\tau=1$) are
relevant. In contrast, for $V_0 > 35$ meV energy levels which come
from both valleys lie in the first Landau gap leading to a more
complex energy spectrum. Focusing on the first Landau gap, at a
fixed $B$ and $\Delta$ there is a critical potential well depth
$V^{v}_{0}$, that defines the crossover from the ``single-valley''
energy spectrum to the ``two-valley'' spectrum. This effect is
rather robust and can be more easily identified when $\Delta$ is
large. Since in the experiments the energy gap is usually fixed we
below focus on the magnetic field dependence of $V^{v}_{0}$.

\begin{figure}[t]
%%% Requires \usepackage{graphicx}
\includegraphics[width=6.2cm, angle=270]{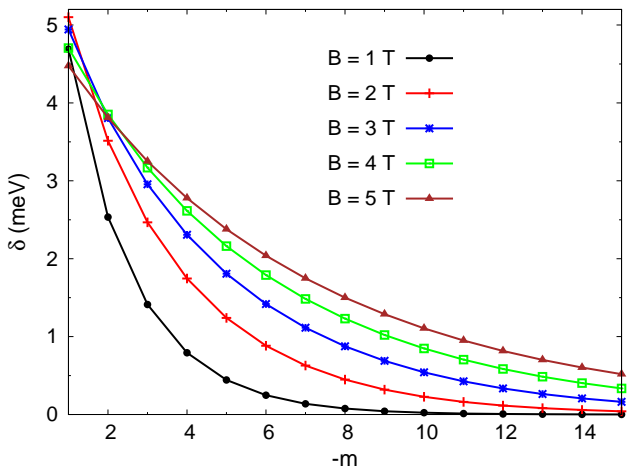}
\caption{Energy spacing $\delta = E(m, \tau=1) - E(m+1, \tau=1)$
between successive energy levels as a function of the angular
momentum number $m$ with $-15\le m\le -1$. The magnetic field is
$B$, the mass term is $\Delta=40$ meV. The potential is
$V_0=V^{g}_{0}$, thus by definition $E(m=0, \tau=1)$ is the lowest
lying energy in the middle of the first Landau gap.}\label{split}
\end{figure}

To explore the energy spectrum in the first Landau gap we first
identify the energy level $E(m=0, \tau=-1, V_0=0) = +\Delta/2$. By
increasing $V_0$ this energy level decreases and we define the
critical potential $V^{v}_{0}$ which satisfies
\begin{equation}\label{crit1}
E(m=0, \tau=-1, V_0=V^{v}_{0}) =-\Delta/2.
\end{equation}
For $V_0>V^{v}_{0}$ energies from both valleys lie in the first
Landau gap and eventually the spectrum becomes rather complex with
many crossing points appearing at random values of $V_0$, $B$ due
to the broken valley degeneracy. The critical potential
$V^{v}_{0}$ can be considered as the maximum allowed value of
$V_0$ which leads to a single-valley energy spectrum in the first
Landau gap. To specify the optimum dot energy range we also need
to determine the minimum value of $V_0$, consequently, we need to
choose a reference ground energy in the first Landau gap. For this
purpose we identify the energy level $E(m=0, \tau=1,
V_0=0)=-\Delta/2$ which decreases with $V_0$, and we define the
critical potential $V^{g}_{0}$ which satisfies
\begin{equation}\label{crit2}
E(m=0, \tau=1, V_0=V^{g}_{0}) =-\Delta/2 - E_L/2.
\end{equation}
with $E_L$ being the value of the first Landau gap as indicated in
Fig.~\ref{spectrum}. This definition of $V^{g}_{0}$ ensures that
the lowest level in the first Landau gap lies in the middle of
this gap. As a result, when the $B$ field is high enough the few
lowest discrete levels are energetically isolated lying far away
from the bulk Landau levels.

Both critical potentials $V^{v}_{0}$ and $V^{g}_{0}$ are magnetic
field dependent. In Fig.~\ref{valley} we plot $V^{v}_{0}$ and
$V^{g}_{0}$ versus the magnetic field for three values of
$\Delta$. The potential $V^{v}_{0}$ decreases as the field
increases and the field dependence is rather large for small
fields when the difference between the mass-induced gap and the
Landau gap is small. In contrast, if the magnetic field is high
enough then $V^{v}_{0}$ varies slowly and eventually $V^{v}_{0}
\rightarrow \Delta$. For example, our calculations show that at
$B=20$ T $V^{v}_{0}\approx 41.5$ meV for $\Delta=40$ meV and
$V^{v}_{0}\approx 20.7$ meV for $\Delta=20$ meV. The potential
$V^{g}_{0}$ has almost a linear field dependence but for low
enough fields ($B<1$ T) the field dependence can be more
complicated. This regime is not of particular interest here
because of the small value of the Landau gap, which in a realistic
sample becomes even smaller due to the level broadening. The
energies coming from the two valleys lie in a different energy
range provided $V^{v}_{0}>V^{g}_{0}$, and according to
Fig.~\ref{valley} this condition is satisfied only for $B \lesssim
1.7$ T when $\Delta=20$ meV. But, when $\Delta=40$ meV
$V^{v}_{0}>V^{g}_{0}$ even when the magnetic field~\cite{note0} is
as high as 6 T. A larger $\Delta$ results in a broader $B$ field
range in which $V^{v}_{0}>V^{g}_{0}$. This feature might be
advantageous to easily separate the energies from the two valleys
and define a dot with a specific valley. However, engineering a
large $\Delta$ cannot be guaranteed and, as quantified below, by
increasing $\Delta$ both the first Landau gap and the energy
spacing between successive levels in the gap decrease.
Consequently, a very large $\Delta$ is not necessarily ideal.

To obtain a better insight into the energy spectra we plot in
Fig.~\ref{split} the energy spacing between successive energy
levels $\delta = E(m, \tau=1) - E(m+1, \tau=1)$ lying in the first
Landau gap at $V_0 = V^{g}_{0}$. All the relevant levels
correspond to $m\le 0$ and originate from the $-\Delta/2$ Landau
level. According to Fig.~\ref{split}, for all magnetic fields the
largest spacing occurs between the two lowest levels that
correspond to $m=0$, $-1$ in order of increasing energy. However,
the spacing, in general, can exhibit a strong $B$ dependence for
the field range considered; as $B$ increases the spacing tends to
increase drastically for large $|m|$ values. For $m=-10$ the
spacing is vanishingly small at $B=1$ T, but is about $1$ meV at
$B=5$ T. This is due to the fact that larger $|m|$ states are
shifted nearer the origin of the well because of the increase of
the magnetic barrier so their energies start to deviate
significantly from the bulk Landau level $-\Delta/2$. The
conclusion is that by tuning the magnetic field the system can be
switched from a few-level dot to a many-level dot with an
appreciable energy spacing. At $B=4$ T the Zeeman splitting is
about $0.46$ meV, for $g=2$ $g$-factor, and $k_{B}T \approx 0.34$
meV at $T=4$ K. These values are much smaller than a typical
spacing of $\delta\approx 2-4$ meV, (Fig.~\ref{split}) allowing
the spectroscopy of the dot levels.~\cite{freitag16} For the
results in Fig.~\ref{split} the effective width of the potential
well is $\sqrt{2}L=40$ nm. Increasing the width to $\sqrt{2}L=80$
nm the $m=0$ energy level lies closer to $-V_0$ and the typical
energy spacing decreases, e.g., at $B=2-4$ T the spacing between
the two lowest levels is about 1.5 meV. The results are only
slightly sensitive to the details of the potential well
profile~\cite{maksym} provided the effective width of the well
remains fixed.

\begin{figure}[t]
%%% Requires \usepackage{graphicx}
\includegraphics[width=6.3cm, angle=270]{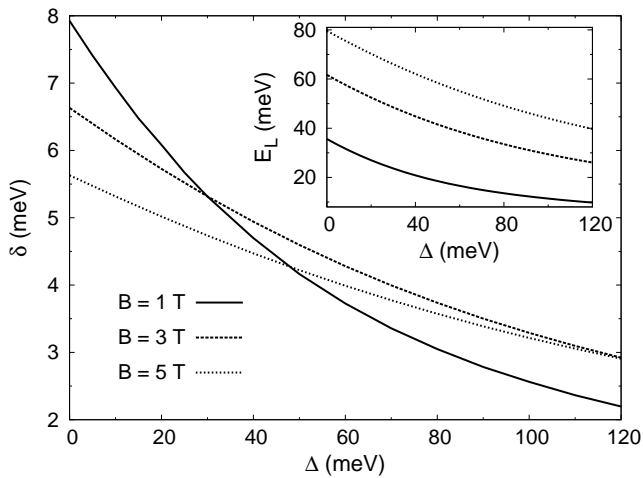}
\caption{Energy spacing $\delta=E(m=-1, \tau=1)-E(m=0, \tau=1)$ as
a function of the mass term $\Delta$ at different fields $B=1$, 3,
5 T. The potential is $V_0=V^{g}_{0}$, thus by definition $E(m=0,
\tau=1)$ lies in the middle of the first Landau gap. The inset
shows the first Landau gap at $B=1$, 3, 5 T.}\label{gap}
\end{figure}

According to the above analysis for the proper $B$, $V_0$ the
discrete energies coming from the two valleys can be adjusted in
different ranges, and in the first Landau gap only $K$-valley
($\tau=1$) energies can exist; with an energy spacing of a few meV
for the low-lying energies. An important issue is how these
properties are affected by the mass term $\Delta$. To explore this
issue we plot in Fig.~\ref{gap} the energy spacing $\delta=E(m=-1,
\tau=1)-E(m=0, \tau=1)$ as a function of $\Delta$ at three
different magnetic fields and $V_0=V^{g}_{0}$. A more general
$\Delta$ and $V_0$ dependence of the energies is presented in
Appendix B. In Fig.~\ref{gap} the condition $V^{v}_{0} >
V^{g}_{0}$ is not necessarily satisfied, and only for $\Delta
\gtrsim$ 10 meV the energy level $E(m=-1, \tau=1)$ corresponds to
the first excited level in the first Landau gap for all the fields
considered. For smaller values of $\Delta$ the level $E(m=0,
\tau=-1)$ is relevant. As seen in Fig.~\ref{gap}, the energy
spacing $\delta$ decreases with $\Delta$ because the effective
mass of the carriers increases, therefore the Dirac system starts
to resemble a Schrodinger one. At $B=1$ T a mass term of
$\Delta\gtrsim16$ meV is needed to induce only $K$-valley energies
in the first Landau gap, i.e., the condition $V^{v}_{0} \gtrsim
V^{g}_{0}$ is satisfied. At $B=3$ T the critical value of $\Delta$
increases to $\Delta\gtrsim26$ meV. Despite this increase the
induced Landau gap at $B=3$ T or 5 T (Fig.~\ref{gap} inset) is
still more than twice larger than at $B=1$ T, while the
corresponding decrease in $\delta$ is small. As a result at higher
magnetic fields the discrete energies of the dot can be put
further away from the bulk Landau levels and still $\delta$ can
have an appreciable value of $\delta \approx 4$ meV for
$\Delta\approx50$ meV.

\begin{figure}[t]
%%% Requires \usepackage{graphicx}
\includegraphics[width=11.5cm, angle=270]{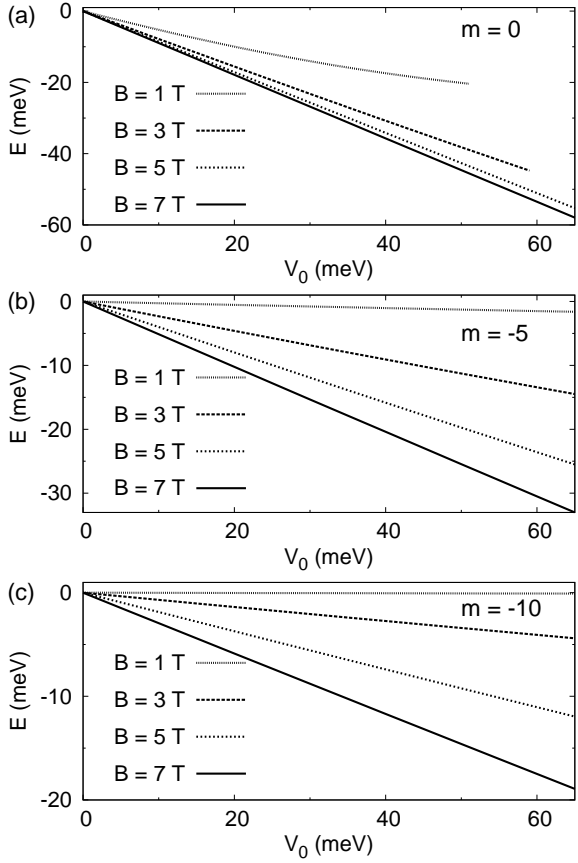}
\caption{Energies as a function of the potential well for
different magnetic fields $B$ and $\tau=1$, $\Delta=40$ meV. All
the levels are shifted by $\Delta/2$; $E\rightarrow E+\Delta/2$,
thus $E=0$ for $V_0=0$. Only energy levels in the first Landau gap
are shown, thus $V_0$ cannot be taken arbitrarily large; (a) at
$B=1$ T and 3 T. The angular momentum number $m$ is indicated in
each frame.}\label{EmB}
\end{figure}

In this work, the lowest energy level of the quantum dot is
defined in the first Landau gap by the $m=0$ level for $\tau=1$.
It is thus useful to explore the variation of this level with
respect to the external parameters $V_0$ and $B$. As seen in
Fig.~\ref{spectrum}(a) for $\Delta=0$ and $B=1$ T this energy
cannot be less than $-V_0$. This feature is always valid and the
energy difference from the bottom of the well is small when the
effective width of the well $L$ is large. Furthermore, as the $B$
field increases the lowest dot energy shifts closer to $-V_0$, and
eventually the field dependence of the energy becomes weak at high
fields when the maximum amplitude of the state occurs in the
quantum well region. Some of these trends are quantified in
Fig.~\ref{EmB}. Specifically, at $V_0=50$ meV the $m=0$ energy
shifts closer to $-V_0$ and changes by over $90\%$ when the field
increases from $B=1$ T to $B=3$ T. At $B=5$ T the $m=0$ state is
almost entirely localised in the quantum well region, therefore,
by further increasing the field to $B=7$ T the corresponding
energy changes by less than $5\%$. On the contrary, focusing on
$m=-5$, $-10$ and tuning the field from $B=5$ T to 7 T the
corresponding energies change by over $28\%$ and $57\%$
respectively. For these values of $m$ small changes in the energy
are observed at much higher magnetic fields.

\begin{figure}[t]
%%% Requires \usepackage{graphicx}
\includegraphics[width=6.2cm, angle=270]{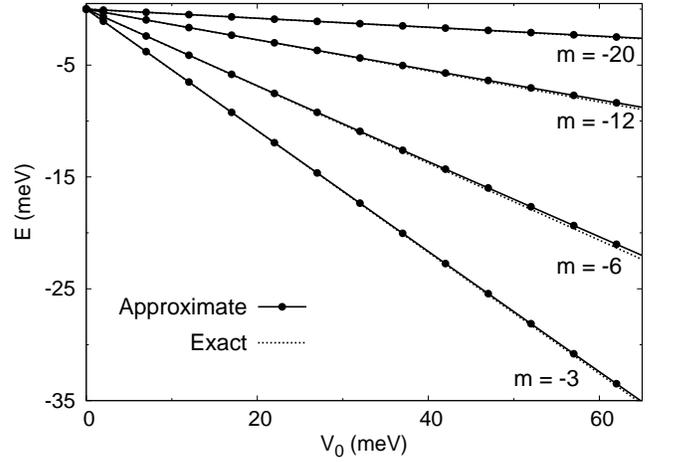}
\caption{Exact and approximate energies as a function of the
potential well at $B=5$ T, $\Delta=0$ and different angular
momentum numbers $m$.}\label{apprE}
\end{figure}

\section{Approximate method}\label{approxi}

Approximate methods for graphene systems are rather rare in the
literature, since the differential equations describing the
systems do not usually allow for any simplified assumptions to be
made. However, if a simple condition is approximately fulfilled by
the two envelope functions $f_i$ then some assumptions can be
made. This idea is followed here.

Specifically, in this section, we develop an approximate method to
determine the $m\le0$ energies $E$ which lie in the first Landau
gap and satisfy the limit $E\rightarrow0$ ($-\Delta/2$) when the
potential $V\rightarrow0$. These energies constitute the dot
energies of interest studied in Sec.~\ref{levels}. The
approximation is particularly good in the regime where one of the
two components $f_i$ dominates. Such a regime occurs when the
potential depth is small, and as specified below the small
potential depth is related to the angular momentum and the
magnetic field.

For the approximate method we start by defining the two functions
$f^{\pm}=f_{1}\pm f_{2}$. Using the equations for $f_i$ derived in
Appendix A [Eqs.~(\ref{f1f2}) and (\ref{f2f1})] it can be easily
shown that $f^{\pm}$ satisfy the two coupled first order
differential equations
%\begin{subequations}
%\begin{eqnarray}
%(V+U)f^{+} + \frac{\tau\Delta}{2} f^{-} - \gamma \frac{df^{-}}{dr} = Ef^{+}, \\
%(V-U)f^{-} + \frac{\tau\Delta}{2} f^{+} + \gamma \frac{df^{+}}{dr}
%= Ef^{-}.
%\end{eqnarray}
%\end{subequations}
\begin{equation}\notag
(V+U)f^{+} + \frac{\tau\Delta}{2} f^{-} - \gamma \frac{df^{-}}{dr}
= Ef^{+},
\end{equation}
\begin{equation}\notag
(V-U)f^{-} + \frac{\tau\Delta}{2} f^{+} + \gamma \frac{df^{+}}{dr}
= Ef^{-}.
\end{equation}
To proceed with the approximate method it is more convenient to
work with the second order equations
\begin{equation}\notag
\frac{d^2f^{+}}{dr^2}+\left(\frac{(E-V)^2}{\gamma^2}-\frac{U^2}{\gamma^2}-\frac{\Delta^2}{4\gamma^2}\right)
f^{+}-\frac{(U-V)'}{\gamma}f^{-}=0,
\end{equation}
\begin{equation}\notag
\frac{d^2f^{-}}{dr^2}+\left(\frac{(E-V)^2}{\gamma^2}-\frac{U^2}{\gamma^2}-\frac{\Delta^2}{4\gamma^2}\right)
f^{-}-\frac{(U+V)'}{\gamma}f^{+}=0,
\end{equation}
with $(\tau\Delta)^2=\Delta^2$. If the two functions $f_{1}$,
$f_{2}$ are localised in identical regions, or with strong
overlap, and a regime can be found satisfying $f_{2}\gg f_{1}$
then $f^{\pm}\approx \pm f_{2}$ and to a good approximation the
two equations above decouple. For $\Delta=0$ this condition is
exact for the zeroth Landau level because when $V=0$ then $f_1=0$
and only $f_2\ne0$. Consequently, if $V\ne 0$ but small enough
then we expect the inequality $f_{2}\gg f_{1}$ to be approximately
satisfied. In this case, the exact second order equations lead to
the following approximate equation for the dominant component
$f_{2}$
\begin{equation}\label{aproxE}
\frac{d^2f_{2}}{dr^2} + K^{2}_{\pm} f_{2}=0,
\end{equation}
and the energy dependent coefficient is
\begin{equation}\label{k2}
K^{2}_{\pm}(r, E) =  \frac{(E-V)^2}{\gamma^2} -
\frac{U^2}{\gamma^2} -\frac{\Delta^2}{4\gamma^2} + \frac{ (U\pm
V)'}{\gamma}.
\end{equation}
The differential Eq.~(\ref{aproxE}) is useful because it has a
Schr\"odinger form. An interesting feature which simplifies the
analysis is that $K^{2}_{\pm}$ has no singular point; unlike the
coefficient $Q^{2}_{i}$ which appears in the exact Eq.~(\ref{Q2})
and at $R_0$ $Q^{2}_{i}(R_0)\rightarrow -\infty$ with
$V_{\mp}(R_0)=0$. Therefore, a confined state $f_2$ that is a
solution to Eq.~(\ref{aproxE}) has an oscillatory amplitude in the
spatial region~\cite{schiff} where $K^{2}_{\pm}>0$, and a decaying
amplitude where $K^{2}_{\pm}<0$. The latter inequality indicates
that for a potential $V$ that rises asymptotically the existence
of a confined state depends on the strength of $V$ and
$A_{\theta}$, in agreement with a previous work.~\cite{giavaras09}

Considering the quantum dot system, $K^{2}_{\pm}$ as a function of
the radial distance $r$ has a positive maximum
($dK^{2}_{\pm}/dr=0$) for all $m$ but zero for which $K^{2}_{\pm}$
diverges at $r=0$. The position of the maximum is sensitive not
only to the dot parameters $B$, $V_0$, $L$, $m$ but also the dot
energy-solution of Eq.~(\ref{aproxE}). When $\Delta=0$ this energy
satisfies the physical requirement $E\rightarrow 0$ for
$V\rightarrow0$, i.e., it converges to the zeroth Landau level.
The two exact equations satisfied by $f^{\pm}$ have also been
applied to potentials $V$ that increase as power laws, and quantum
states with large positive values of $m$. Then to a very good
approximation~\cite{giavarasE} $f_{1}\approx \pm f_{2}$. This
regime is very different from the one examined here resulting in a
markedly different energy spectrum.

The solution of Eq.~(\ref{aproxE}) is only slightly sensitive to
the sign of the $V'$ term, provided $(U \pm V)' \approx U'$ in the
region where $f_{2}$ is localised. This condition can, for
example, be satisfied when $B$ is high and simultaneously $V$ is
small. The term $(U \pm V)'$ however cannot be completely
discarded because it can lead to $K^{2}_{\pm}<0$ for all $r$
failing to predict the dot energy. For clarity we consider only
$(U-V)'$, and to quantify the approximate results we plot in
Fig.~\ref{apprE} some approximate energy levels together with the
exact ones as a function of the potential depth $V_0$. The
agreement is very good and the correct $V_0$ dependence of the
energies is predicted.

\begin{figure}%%%%[t]
%%% Requires \usepackage{graphicx}
\includegraphics[width=6.3cm, angle=270]{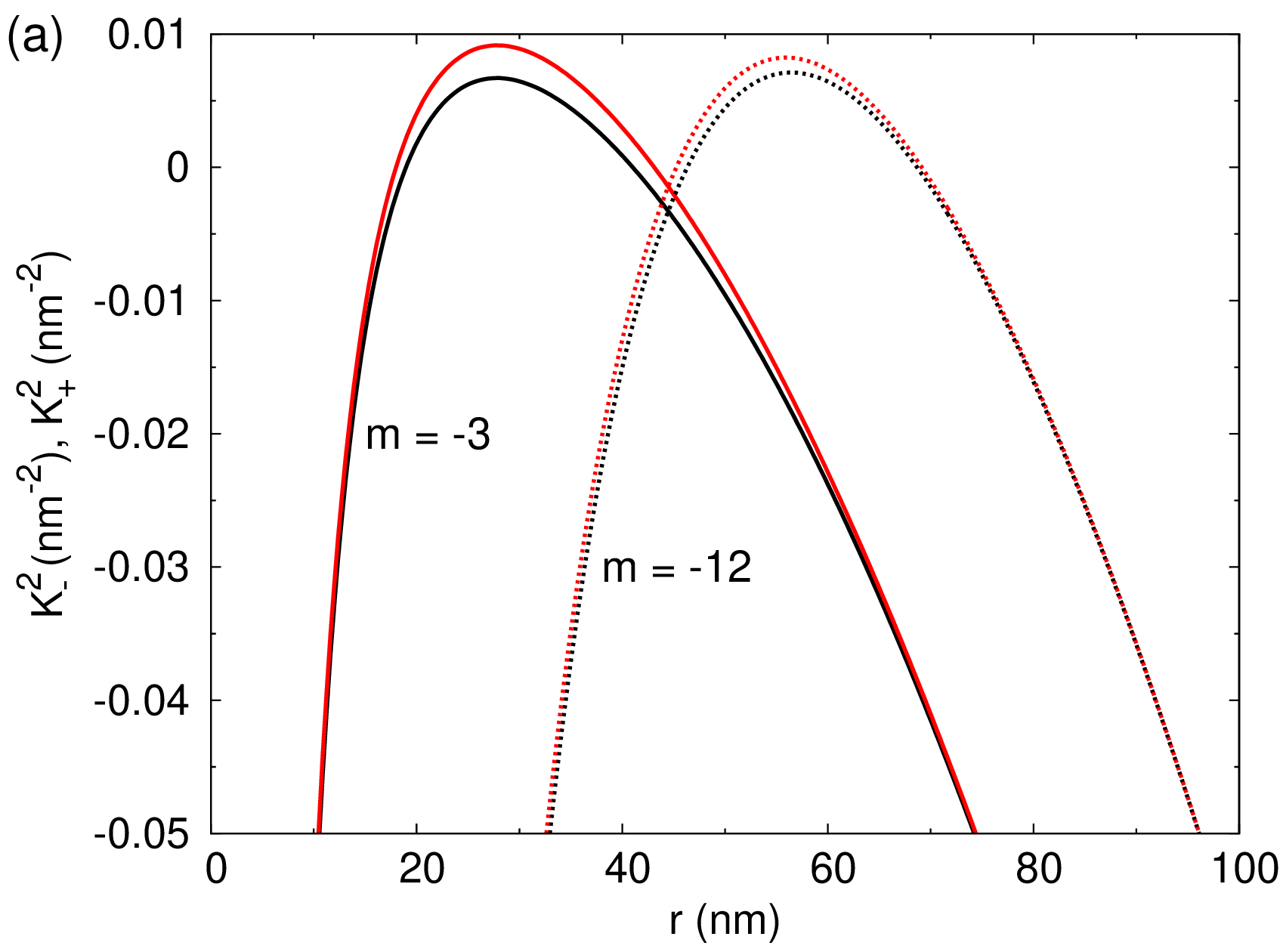}
\includegraphics[width=8.cm, angle=270]{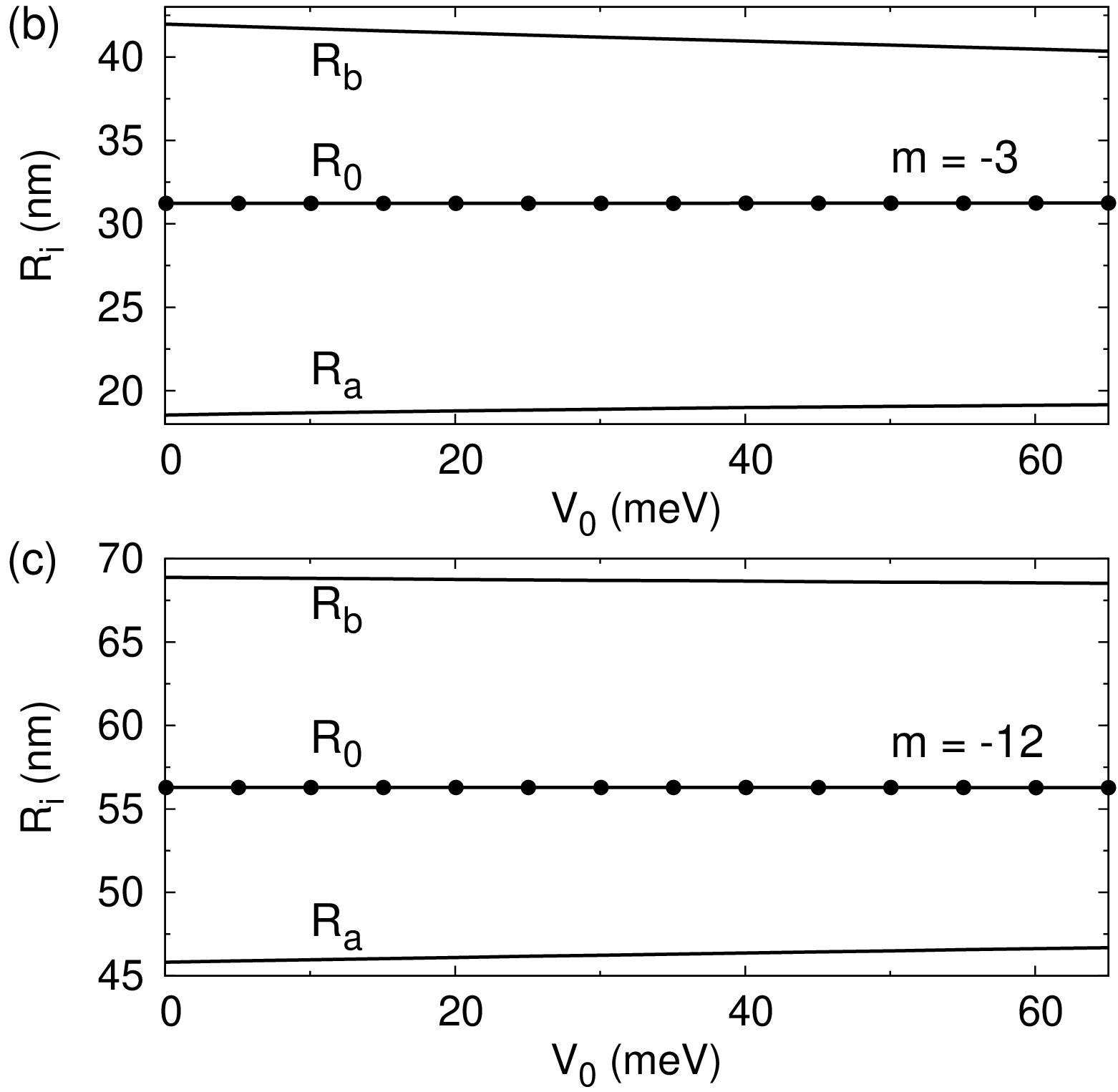}
\caption{(a) Energy dependent coefficients $K^{2}_{-}$,
$K^{2}_{+}$ defined in Eq.~(\ref{k2}) in the main text, as a
function of the radial coordinate for $B=5$ T, $\Delta=0$,
$V_0=37$ meV and two values of $m$. The curves shown in black
(red) correspond to $K^{2}_{-}$ ($K^{2}_{+}$). [(b), (c)] Turning
points $R_{a}$, $R_{b}$ satisfying
$K^{2}_{-}(R_a)=K^{2}_{-}(R_b)=0$, and singular point $R_0$
satisfying $E=V(R_0)$ for $m=-3$ and $m=-12$ respectively. On the
length scale shown the variation of $R_0$ is vanishingly
small.}\label{clas}
\end{figure}

The approximation is somewhat better for larger negative values of
$m$, but when the absolute energy is very small numerical errors
in the computations can change this trend. The error~\cite{note}
for the approximate energies plotted in Fig.~\ref{apprE} is about
$1-2\%$, and even better agreement is achieved for specific values
of $V_0$ and $m$. From Fig.~\ref{apprE} we can also extract that
the error slightly increases with $V_0$, and specifically, in the
chosen $V_0$ range the increase is about $1\%$. All these trends
are consistent with the fact that as $V_0$ increases the component
$f_{1}$ acquires a larger amplitude so the basic assumption
$f_{2}\gg f_{1}$ becomes gradually less accurate especially when
$|m|$ is small. The agreement with the exact numerical energies is
better when the effective width $L$ of the quantum well increases,
and as a result the term $V'$ decreases in the well region. For
$\sqrt{2}L=80$ nm the maximum error for the same parameter values
shown in Fig.~\ref{apprE} is less than $1\%$, and small errors of
the order of $0.1\%$ can be achieved. The key idea behind the
approximation is that the two components $f_i$ are localised in
nearly the same regions but one of the two components dominates.
The latter condition is well satisfied at high $B$, small $V_0$
and large $L$.

The position of the singular point that appears in the exact
Eq.~(\ref{Q2}) but not in the approximate Eq.~(\ref{k2}) deserves
some further investigation. The fact that there is no singular
point in the approximate Eq.~(\ref{k2}) does not necessarily imply
that the singular point is located (far) away from the region
where $f_2$ is localised, and can be ignored. To quantify this
argument we plot in Fig.~\ref{clas}(a) the coefficient
$K^{2}_{\pm}(r)$ versus the radial coordinate for the parameters
$B=5$ T, $\Delta=0$ and $V_0=37$ meV. The function $f_2$ is
localised in the region where $K^{2}_{\pm}(r)>0$, and this region
shifts away from the origin for larger values of $m$, in agreement
with the behaviour exhibited by the Landau states. Because
$K^2_{+}(r)$ is not very different from $K^{2}_{-}(r)$, we
consider for simplicity only $K^{2}_{-}(r)$ and define the two
``turning'' points $R_a$, $R_b$ with
$K^{2}_{-}(R_a)=K^{2}_{-}(R_b)=0$, $R_b>R_a$. In
Figs.~\ref{clas}(b) and (c) we plot $R_{a}$, $R_{b}$ together with
the singular point $R_0$ of Eq.~(\ref{Q2}) satisfying $E=V(R_0)$.
For each value of $V_0$ the condition $R_a<R_0<R_b$ is satisfied,
indicating that $R_0$ occurs within the region where $f_2$ is
localised and has large amplitude. The conclusion from the two
examples illustrated in Figs.~\ref{clas}(b) and (c) is that the
approximate method is applicable to the regime where $R_0$ cannot
be ignored in the exact Eq.~(\ref{Q2}). The occurrence of $R_0$ is
needed to give rise to a non-zero $f_{1}$, when $V_0\ne0$, in
approximately the same region as that where $f_2$ is localised.
The fact that $R_0$ depends only weakly on $V_0$ [Figs.~7(b) and
7(c)] indicates that some further approximations can be made.
Investigation of the numerically exact values of $R_0$ shows that
$R_0\approx R$ for some values of $m$, where
$R=\sqrt{(2|m|+1)\hbar/eB}$ is the peak position of $f_2$ when
$V_0=0$. Then $E$ can be determined analytically, $E\approx V(R)$,
but quantifying this approximation is beyond the scope of this
paper.

Finally, based on the assumptions behind the approximate method,
it can be easily demonstrated that the method can be applied
equally well to quantum dot potentials which continuously
increase, e.g., $V(r)=V_g r^n$, with $n>0$ and $V_g>0$. However,
in this case the physics is different from a potential that is
asymptotically constant. The reason is that the energy deviation
from the zeroth Landau level increases with $|m|$, as the Landau
states which are localised away from the origin are affected more
by the potential $V(r)$. For completeness one example for this
case is presented in Appendix C.

\section{Conclusions}\label{conclu}

In summary, we considered a graphene sheet in a perpendicular
magnetic field and explored the energy spectrum of a quantum dot
formed in the first Landau gap. The discrete levels of the quantum
dot emerge from the Landau levels using a potential well which can
be experimentally realised and efficiently tuned with an STM
tip.~\cite{freitag16, freitag18}

In our graphene system the valley degeneracy is broken and as a
result the dot energy spectrum can be rather complex exhibiting
many irregular level crossings and small energy spacings which are
sensitive to the applied fields. However, as demonstrated in this
work in the first Landau gap and for the appropriate potential
well, magnetic field, and mass-induced gap the dot energy spectrum
can have a simple pattern with discrete energies coming from one
of the two valleys only. In this part of the spectrum there are no
energy crossings and the lowest energy level corresponds to $m=0$
angular momentum, while $m$ successively decreases by $-1$ for
each higher energy level.

The magnetic field dependence of the energy levels and the effect
of the mass term were examined. It was shown that the discrete
energies of the dot in the first Landau gap lie away from the bulk
Landau levels and the typical energy spacing can be large enough,
of a few meV, when the magnetic field is $3-5$ T, and the
mass-induced gap is about 50 meV. It was demonstrated that in the
regime where one of the two components of the Dirac state
dominates an approximate method can be developed. Within this
method a Schr\"odinger equation was derived which can predict the
region where the dominant component is localised. The approximate
energies for states with negative angular momentum exhibit the
correct general trends and are in a good agreement with the exact
energies.

The graphene system studied here is experimentally realizable and
the simple energy patterns that were identified arise in a
realistic range of fields. The results of this work could guide
further investigations of confined states in two dimensional
materials.

\setcounter{secnumdepth}{0}   %% no numbering

\setcounter{secnumdepth}{1}

\appendix

\section{Quantum dot equations}

The equations describing a graphene quantum dot in the continuum
approximation have been derived in various previous
works.~\cite{giavaras12, giavaras11} In brief, the eigenvalue
problem defined in Eq.~(1) in the main text can be reduced to two
equations for the radial functions $f_{i}$. This is done with the
substitution~\cite{giavaras12, giavaras11} $\Psi(r, \theta) =
(f_{1}(r)\exp[i(m-1)\theta]$, $if_{2}(r)\exp[i m\theta])/\sqrt{r}$
which leads to
\begin{subequations}\label{f1f2}
\begin{eqnarray}
(V+\tau\Delta/2)f_{1} +\left(U+\gamma\frac{d}{dr}\right)f_{2} = E f_1, \label{f1f2} \\
\left(U-\gamma\frac{d}{dr}\right)f_{1}+(V-\tau\Delta/2)f_{2}=
Ef_2, \label{f2f1}
\end{eqnarray}
\end{subequations}
where
\begin{equation}\notag
U=\gamma\frac{2m-1}{2r}+\gamma\frac{eA_{\theta}}{\hbar},
\end{equation}
includes the terms due to the angular motion and the magnetic
vector potential. For convenience we set
$V_{\pm}=V-E\pm\tau\Delta/2$ and use these two equations to derive
the decoupled equations for each radial envelope function $f_i$
\begin{subequations}
\begin{eqnarray}
\frac{d^2f_1}{dr^2}-\frac{V^{'}}{V_-} \frac{df_1}{dr} +
\left(-\frac{U^2}{\gamma^2}-\frac{U^{'}}{\gamma}+\frac{U}{\gamma}\frac{V^{'}}{V_-}
+ \frac{V_-V_+}{\gamma^2} \right)f_1=0,\notag \\
\frac{d^2f_2}{dr^2}-\frac{V^{'}}{V_+} \frac{df_2}{dr} +
\left(-\frac{U^2}{\gamma^2}+\frac{U^{'}}{\gamma}-\frac{U}{\gamma}\frac{V^{'}}{V_+}
+ \frac{V_-V_+}{\gamma^2} \right)f_2=0, \notag
\end{eqnarray}
\end{subequations}
where prime denotes differentiation with respect to $r$. These two
equations have the general compact form
\begin{equation}\notag
\frac{d^2f_i}{dr^2} + a_i \frac{df_i}{dr} + b_i f_i=0,
\end{equation}
with $i=1$, 2 and the coefficients $a_i$, $b_i$ can be inferred.
We assume~\cite{giavaras11} $f_i = g_i y_i$ and derive that
\begin{equation}\notag
g_i \frac{d^2y_i}{dr^2} + (2 g^{'}_i + a_i g_i  ) \frac{d y_i}{dr}
+ (g^{''}_{i} + a_i g^{'}_{i} + b_i g_i)y_i=0.
\end{equation}
We choose $g^{'}_i/g_i=-a_i/2$ to derive the final equation for
$y_{i}$:  $d^2y_i/dr^2 + Q^{2}_{i} y_i = 0$. The coefficient
$Q^{2}_{i}(r, E)$ is given in Eq.~(\ref{Q2}) in the main text.

\section{Effect of mass term on dot levels}

\begin{figure}%%%%[t]
%%% Requires \usepackage{graphicx}
\includegraphics[width=11.cm, angle=270]{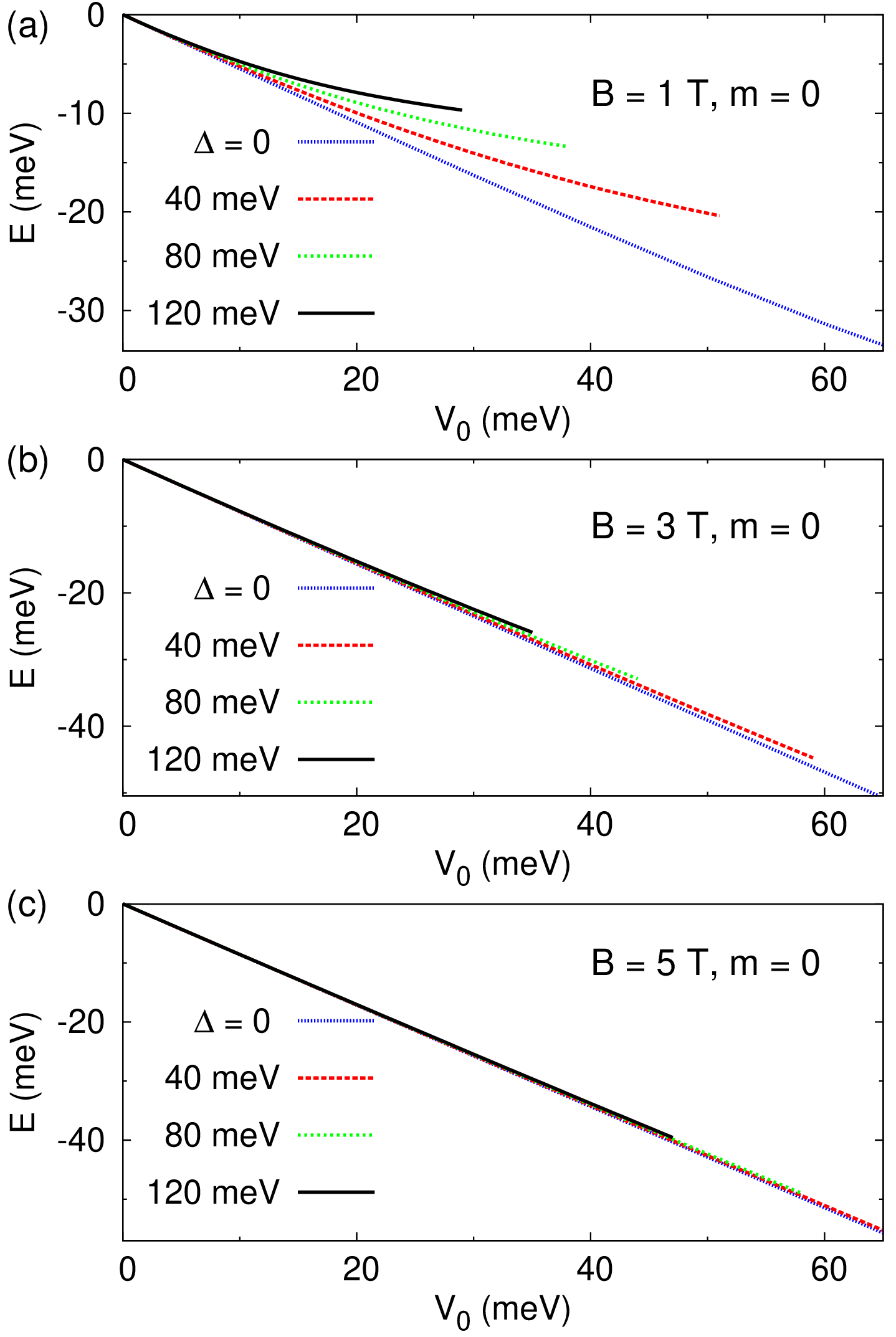}
\includegraphics[width=11.cm, angle=270]{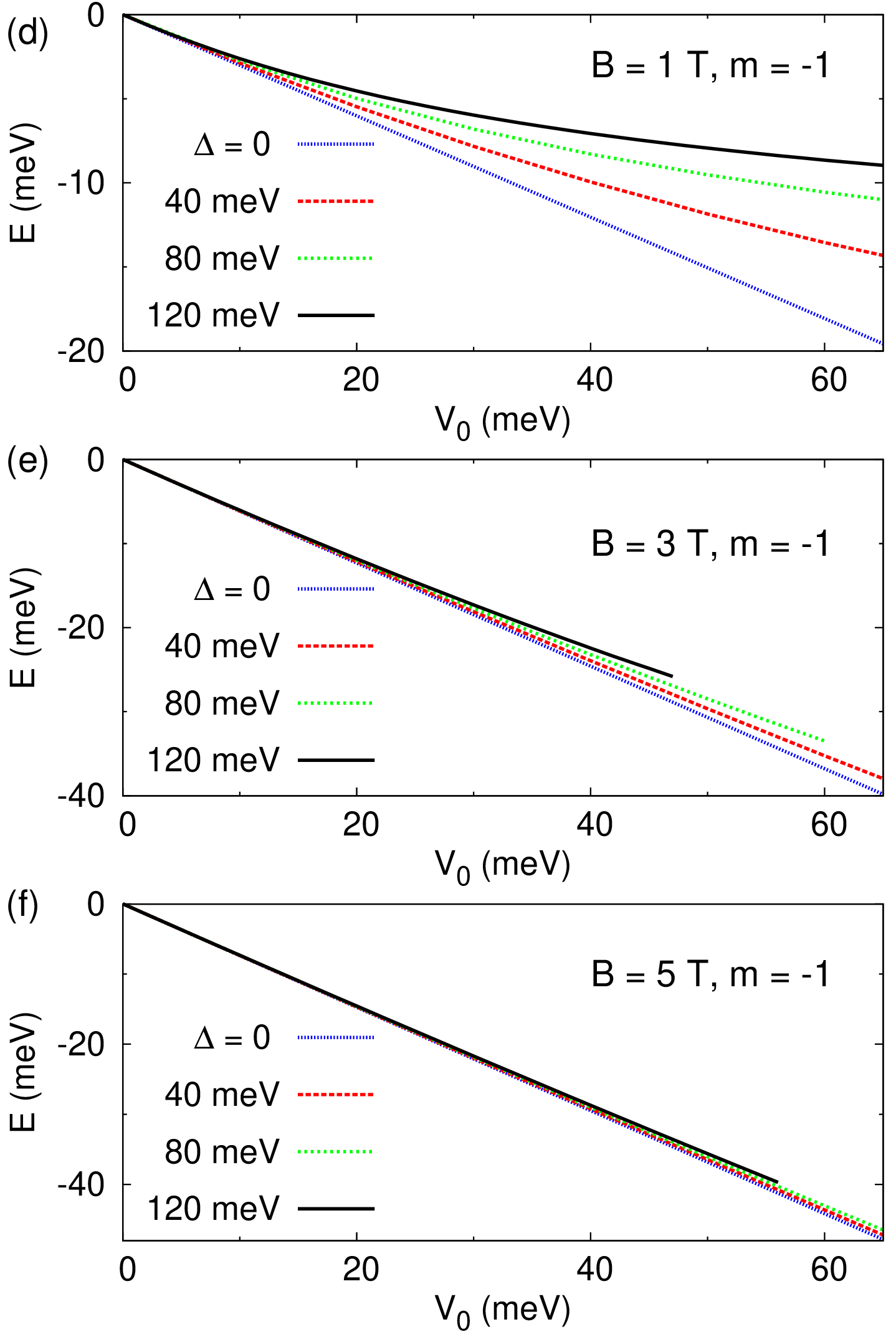}
\caption{Energy levels as a function of the potential well for
different mass terms $\Delta$ and $\tau=1$. For a better
comparison all the levels are shifted by $\Delta/2$; $E\rightarrow
E+\Delta/2$, thus $E=0$ for $V_0=0$. Only energy levels in the
first Landau gap are shown, thus $V_0$ cannot be taken arbitrarily
large. The angular momentum number $m$ and the $B$ field are
indicated in each frame.}\label{m0}
\end{figure}

In the main text, it was shown that in the appropriate range of
parameters the $m=0$, $-1$ energy levels for $\tau=1$ can define
the two lowest levels of the quantum dot in the first Landau gap.
Figure~\ref{m0} shows the effect of the mass term $\Delta$ on
these levels for different magnetic fields. At low magnetic fields
the value of $\Delta$ is important, especially at large $V_0$
values which approach the Landau gap. However, as the field
increases the two levels exhibit a smaller $\Delta$ dependence,
until eventually a shift of around $\Delta/2$ is observed. These
trends are in agreement with the change in the corresponding
energy spacing versus $\Delta$ shown in Fig.~\ref{gap} in the main
text.

\begin{figure}[t]
%%% Requires \usepackage{graphicx}
\includegraphics[width=6.cm, angle=270]{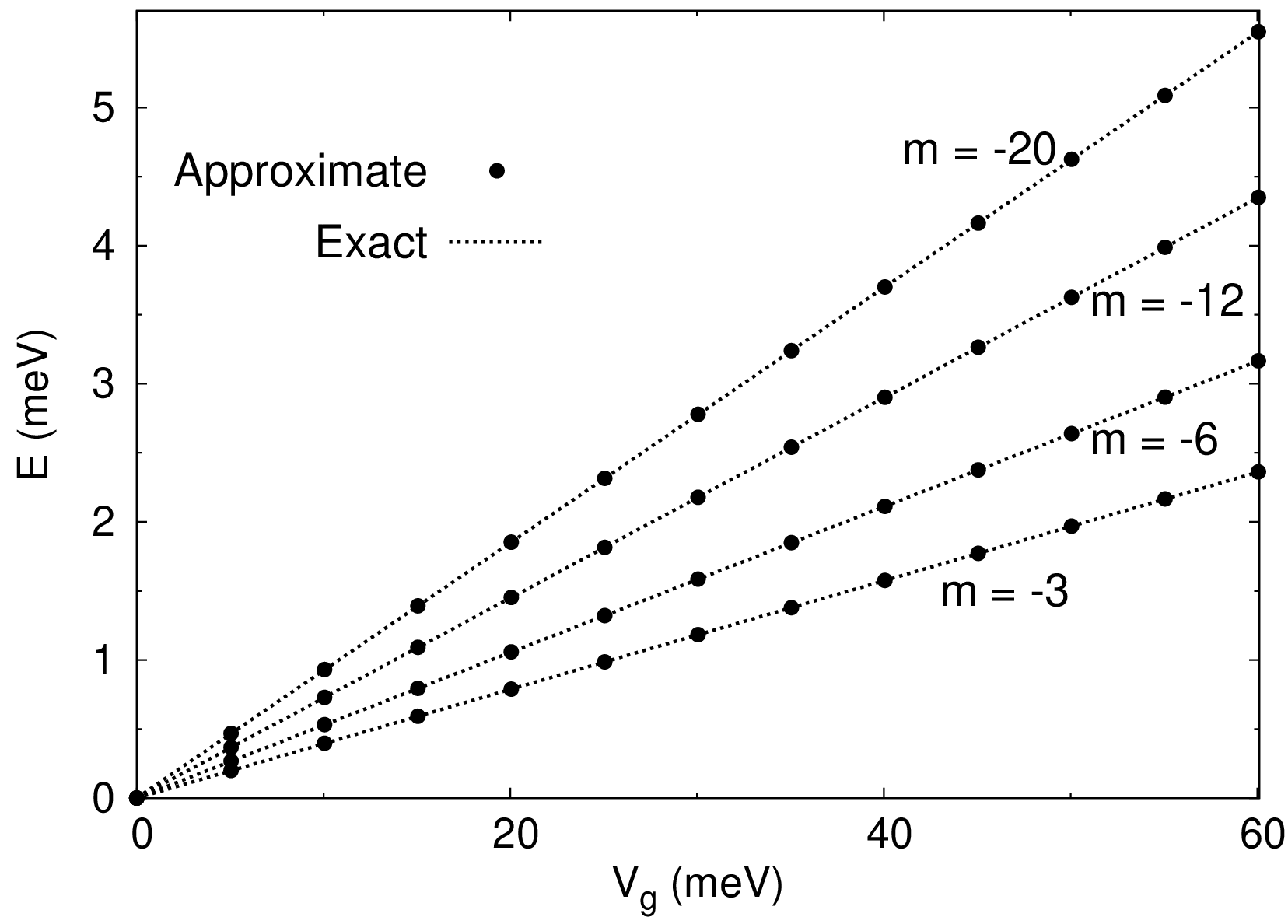}
\caption{Exact and approximate energies as a function of the
potential well at $B=5$ T, $\Delta=0$ and different angular
momentum numbers $m$. The quantum dot potential is $V(r)=V_g r/x$,
with $x=800$ nm.}\label{lin}
\end{figure}

\section{Model quantum dot potential}

To demonstrate the efficacy of the approximate method described in
the main text, we here study a model quantum dot potential. This
is defined by the potential well $V(r)=V_g r/x$, with $x=800$ nm.
Figure~\ref{lin} illustrates the exact and approximate energies as
a function of $V_g$ for different angular momentum $m$. The
overall agreement is excellent and the error is less than $0.1\%$;
depending on the parameters it can be even an order of magnitude
smaller. Similar to Figs.~\ref{clas}(b) and (c) in the main text,
the condition $R_a<R_0<R_b$ is again satisfied. This demonstrates
the importance of the singular point $R_0$ when $V_g\ne 0$.


\begin{thebibliography}{10}
\bibitem{note}
For $m=0$ the error is about $2\%$, whereas when $\sqrt{2}L=80$ nm
the error is less than $1\%$.
\end{thebibliography}
\end{document}